%% file: main_arxiv.tex
\documentclass[aps,prl,reprint,groupedaddress,amsmath,amssymb]{revtex4-2}
\preprint{APS/123-QED}
\usepackage{graphicx}
\usepackage{amsfonts}

\begin{document}
\title{Thickness-Driven Transitions Between Novel Magnetic States in Ferromagnetic Films}
\author{J. J. Mankenberg}
\email[]{m0908394@tamu.edu}
\author{Ar. Abanov}
\affiliation{Department of Physics, Texas A\&M University, College Station, Texas 77843-4242, USA}

\date{\today}

\begin{abstract}
Magnetic materials hosting stable topological spin textures have demonstrated energy efficiency and potential as information carriers in novel logic and memory devices, offering an alternative to magnetic tunnel junction technology. While these structures are well understood in 2D, in 3D their stability, interactions, and topological transitions require further exploration. Here we present two thermodynamically stable topological states, termed the ``hourglass" and ``dome", in centrosymmetric magnetic films of varying thickness. Crucially, we observe thickness-dependent transitions between the two states, with regions of metastability where both configurations coexist. We construct a phase diagram detailing the parameter space for their existence and transitions, provide an effective description of the interactions that mediate the transition, and discuss the implications of this type of state switching.
\end{abstract}

\maketitle

In recent years, magnetic skyrmions have been extensively studied experimentally, numerically and analytically in a multitude of systems; however, their primary arena, is in two condensed matter systems \cite{everschor_2018,tokura_kanazawa_2020,sharma_mishra_2022,nagaosa_tokura_2013}: thin films \cite{yu_2010,heinze_von_2011} and bulk helimagnets \cite{muhlbauer_2009}. In thin films, they are usually stabilized by a local inversion symmetry breaking Dzyaloshinskii-Moriya interaction (DMI)\cite{dzyaloshinsky_1958,moriya_1960}. In bulk, they span the thickness of the material and are often stabilized by an external AC magnetic field\cite{everschor_2011} or bulk DMI in non-centrosymmetric materials. They attract significant research interest for their topological protection, which makes them resistant to noise, defects, and current-induced annihilation, enabling their use as stable information carriers, in high-density memory storage, and logic devices.\cite{finocchio_2016,kang_2016,wiesendanger_2016,garst_2017,jiang_2017,fert_2017,leonov_2016}. 

Despite the numerous benefits of skyrmion based devices, there are still technical obstacles to overcome to make them viable products. Among these is the difficulty of consistently manufacturing the ultrathin films that can store, transport, and read the structures in a useful manner \cite{finocchio_2016,kang_2016,jonietz__2010}. Reproducing desirable skyrmion properties in cheaper, thick films could remedy this difficulty. In addition, utilizing full 3D materials may lead to the higher magnetic and logical connection densities in devices. This is the practical motivation for this work.

In this letter, we introduce two novel topological states in quasi-3D ferromagnetic systems that retain inversion symmetry and are, therefore, absent of DMI. Despite sharing identical boundary conditions, these states represent fundamentally distinct equilibrium solutions to the micromagnetic equations, each localized in three dimensions. We present a phase diagram outlining the conditions for their stability and an interaction model that describes the underlying mechanism driving these transitions. The scope of this work is limited to static properties as dynamics, such as electric current induced motion, breaking, and switching, will be studied in future work.

\begin{figure*}[ht]
\includegraphics[width=\textwidth]{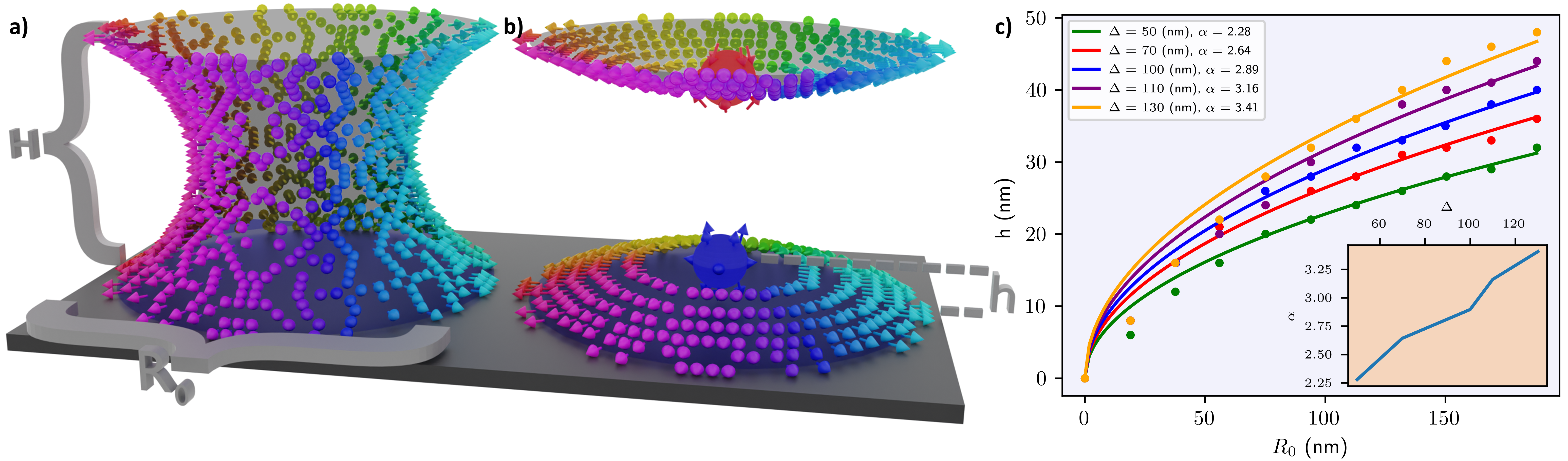}%
\caption{\label{fig:HourglassDome} A side by side view of the two structures stabilized from the same boundary conditions dubbed (a) the hourglass solution and (b) the dome solution. The vectors are drawn from raw data from micromagnetic simulations and represent the in-plane ($m_z = 0$) component of the magnetization. The color scheme depicts local azimuthal angle of the vectors. At a critical film height, such as the one shown here, both configurations have the same total energy and may transition between each other. (c) Equilibrium height $h$ of a single dome versus the fixed boundary radius $R_0$ for a range of domain wall widths $\Delta$. A square root least-squares fit $h = \alpha\sqrt{R_0}$ is plotted along with the data. The inset shows that the best fit values for $\alpha$ increase with $\Delta$. The relation lets $R_0$ determine the size of the structures.}
\end{figure*}

We start by considering thick films composed of interacting $2D$ ferromagnetic layers stacked along the $z$ direction. We characterize the films as thick, as there are no restrictions on their thickness; it is treated as variable and unbounded. The inter-layer exchange coupling is $J_z$ while the intra-layer exchange coupling is $J$ with an easy-axis anisotropy $\lambda $ in the $z$ direction, perpendicular to the film. The Hamiltonian in the continuum limit then reads
\begin{eqnarray}\label{eq:Ham}
\mathcal{H}=\int \text{d}x\text{d}y\text{d}z \Big(&&-\frac{J}{2}\big((\partial_{x}\mathbf{m})^{2}+ (\partial_{y}\mathbf{m})^{2}\big)\nonumber\\
&&-\frac{J_{z}}{2}(\partial_{z}\mathbf{m})^{2} -\lambda m_{z}^{2}\Big)  
\end{eqnarray}
with the algebraic homogeneity condition on the magnetization, $\mathbf{m}^2 = 1$. There is no DMI in the bulk of the film, but there is some form of inversion symmetry breaking at the film interface. Conventionally, this is done by having the film sit on a heavy metal substrate that induces a DMI at the interface. Prominent examples are Pt/Co/Ir multilayers \cite{Maccariello_2018} or hexagonal Fe films on an Ir(111) surface \cite{heinze_von_2011}.

We use a combination of micromagnetic numerical simulations produced with the ubiquitous software packages OOMMF \cite{donahue_porter_1999} and MuMax3 \cite{mumax} to minimize the Hamiltonian in Eq.~\eqref{eq:Ham} with the boundary condition consisting of two opposing skyrmions on either side of the film. The boundary skyrmions are sufficiently stabilized by interfacial DMI to prevent collapse during minimization. They are treated as fixed with a radius $R_0$.

For the purposes of reproducibility, we primarily state tested parameters that fit with experimentally observed skyrmions in CoCrPt films \cite{zheng_wang_ng_2002,roy_nuhfer_laughlin_2003,navas_nam_velazquez_ross_2010}. While we expect these values to change with greater film thickness, we have simulated a wide range of material values, and the results presented here are consistent across this broader parameter space. The size of the simulated mesh is $500$nm wide (x-direction), $500$nm long (y-direction) and a variable height with a mesh cell size of one cubic nanometer. The material parameters are the saturation magnetization M$_s = 0.3\times10^6$A/m, exchange coupling J$ = 2.0\times10^{-11}$J/m, and uniaxial anisotropy $\lambda = 6.2\times10^4$J/m$^3$ acting in the z direction. In addition, to improve numerical stability and prevent an asymmetric structure from forming, an artificial pinning center was created by locally increasing the anisotropy to $1.0\times10^5$J/m$^3$ in a small region at the interface.

There are two sets of initial magnetization configurations that span the same range of film thicknesses. In the first set, two interfacial skyrmions were stabilized at opposing sides of the film boundary. The bulk of the film was set as a uniform $\mathbf{m} = (0,0,1)$ which corresponds to the minimum energy configuration determined by anisotropy. The second set had the boundary skyrmions connected by a uniform skyrmion string. We emphasize that both sets have the same boundary conditions, namely fixed skyrmions of radius $R_0$, and the same range of film thicknesses.

All simulations relaxed to the same two classes of objects shown in Fig.~\ref{fig:HourglassDome}(a) and \ref{fig:HourglassDome}(b) which we refer to as the hourglass and dome solutions, respectively. They exhibit some convexity in their shape which is characterized by the $m_z = 0$ isosurface. This is in sharp contrast to similar solutions in chiral magnets such as skyrmion tubes and chiral bobbers \cite{Rybakov_2015} which are straight and concave respectively. We elaborate on these profiles in the Supplemental Material.

The hourglass is absent of singular points and propagates through the thickness of the film, resembling a skyrmion tube but with a thinning radius at its midpoint. This shape gives it its name. Imposing local twisting of the boundary skyrmions does not change its profile. The dome on the other hand terminates with a Bloch point, similar to a chiral bobber. This dome height $h$, i.e. the distance from the boundary to the Bloch point, is determined by material parameters and the size of the fixed boundary skyrmion $R_0$. It is best described by the relation $h = \alpha\sqrt{R_0}$ where $\alpha$ incorporates the competing micromagnetic energies. This relation breaks down however for small domes about the size of the domain wall width $\Delta = \sqrt{2J/\lambda}$. This is shown alongside numerical data in Fig.~\ref{fig:HourglassDome}(c).

\begin{figure}[ht]
\includegraphics{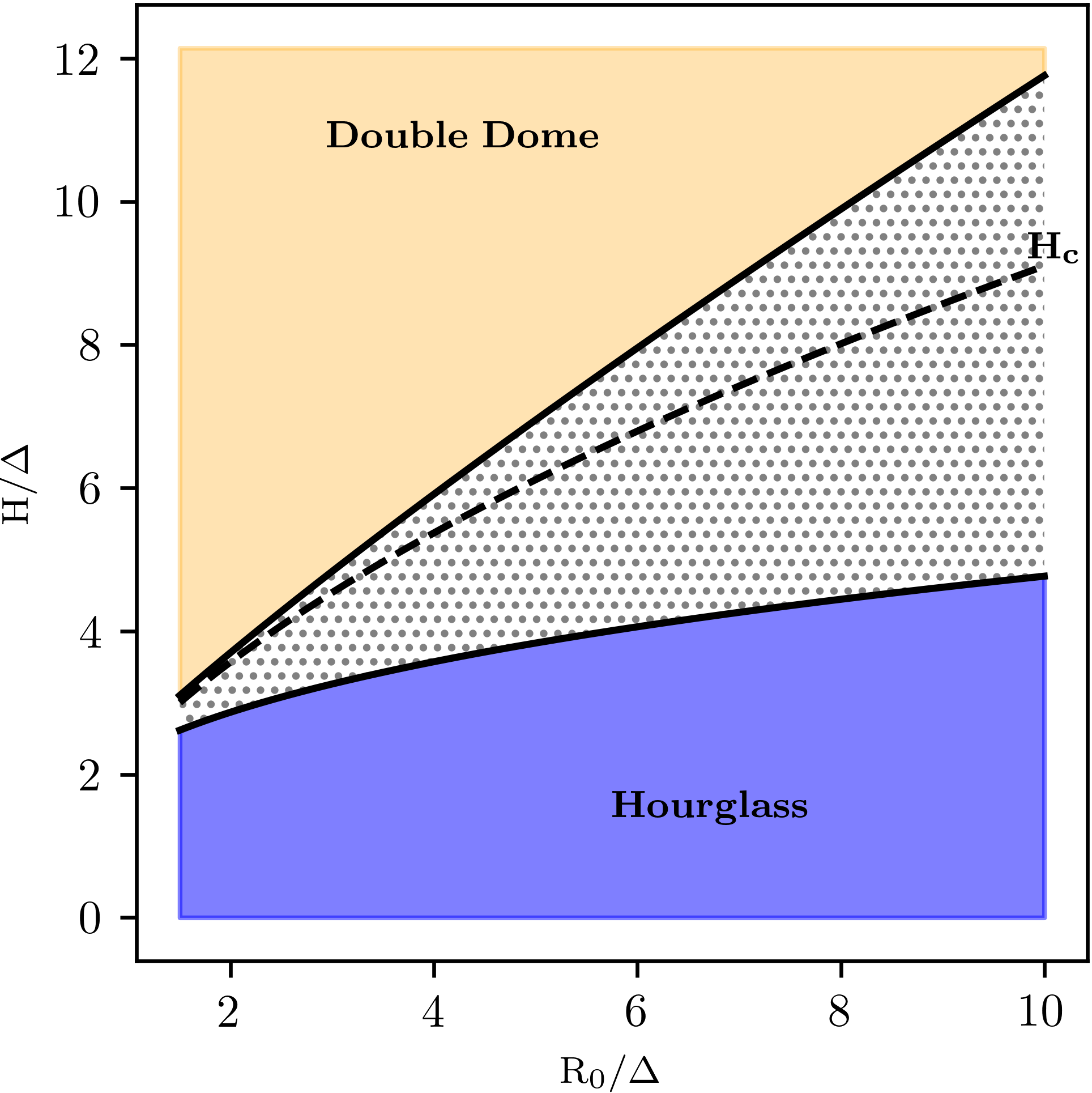}%
\caption{\label{fig:phaseDiagram} Phase diagram of the ground state for isotropic film of varying thickness and boundary skyrmion radius $R_0$. The dotted area within the solid lines is the region of pair metastability, where both the hourglass and double domes can be found. The dashed line corresponds to the critical height and radius at which both solutions have equal energy.}
\end{figure}

\begin{figure*}[ht]
\includegraphics[width=\textwidth]{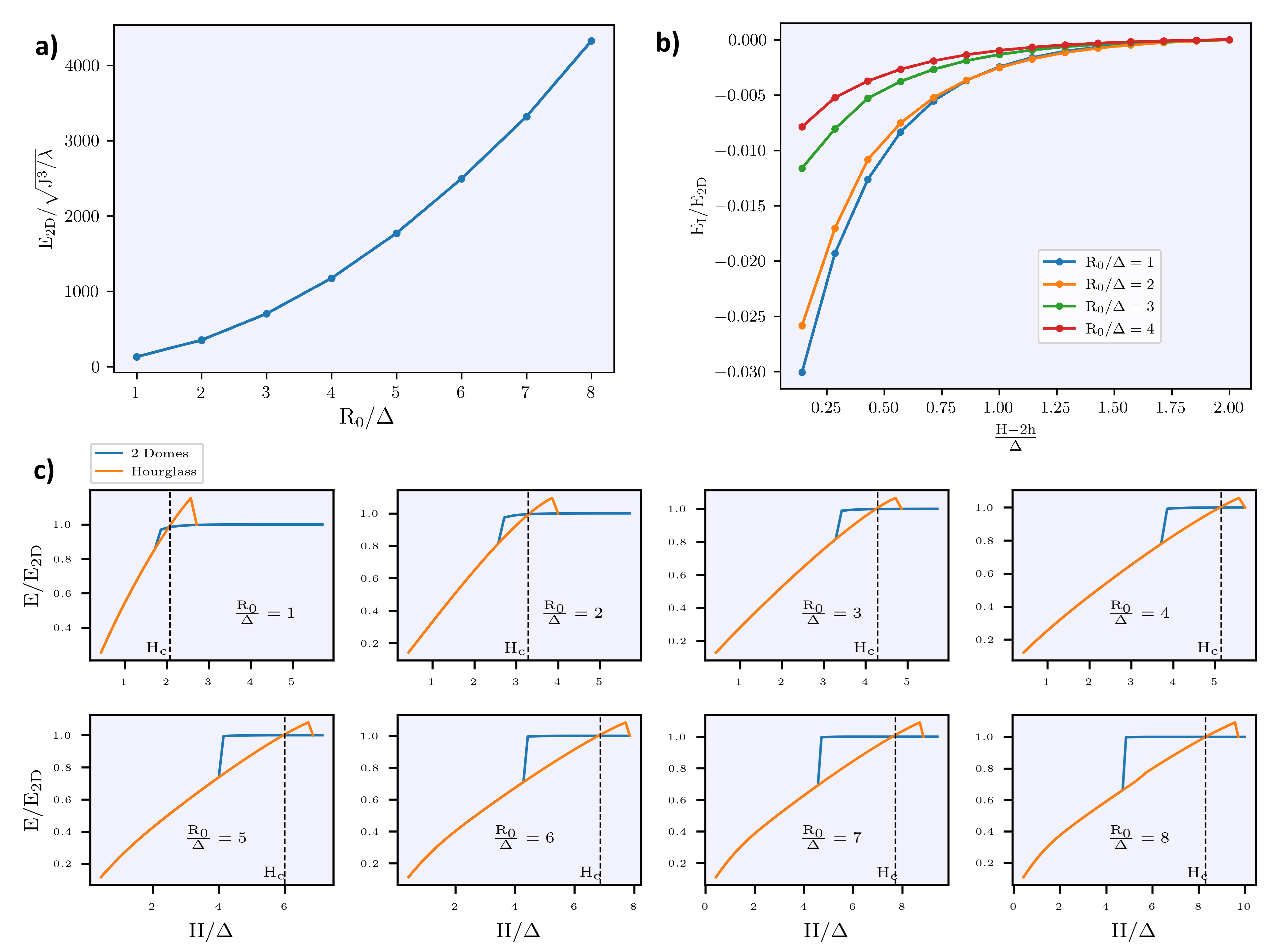}%
\caption{\label{fig:Comp_energylandscape} a) Dimensionless energies of the noninteracting two dome solutions plotted verses the respective dimensionless boundary radius $R_0$. As the domes get bigger the energy increases as a second order power law. b) Detail of interaction energies of dome pair Bloch points for different sized domes versus the separation distance between them. The energies are normalized by the respective energies of two domes at large separation. The profile closely follows a Yukawa potential with a length scale set by domain wall width $\Delta$ as modeled in Eqs.~(\ref{eq:Yukawa}) and (\ref{eq:Hc}). Within a separation of $2\Delta$, the interaction is virtually extinguished by the anisotropy. The strength of the interaction is shown to decrease with dome size. c) The dimensionless energies of the hourglass-dome transition as a function of film thickness (in number of lattice spacing) for a range of base radii $R_0$ taken directly from simulations. The orange and blue curves are the relaxed values from the two sets of initial magnetization.
}
\end{figure*}

To identify the range of parameters corresponding to stable configurations, we calculated the phase diagram for these states and present it in Fig.~\ref{fig:phaseDiagram}. The solid lines represent the hard transition between a pair of opposing domes and the hourglass for both starting magnetization configurations. The middle region corresponds to metastability where both configurations represent local minima in the energy landscape. In this region, reversible transitions between the two configurations may be induced by relatively weak currents, magnetic fields and thermal fluctuations at the same film thickness. The dashed line denotes equal energies between both solutions occurring along a critical height curve.

Despite the lack of Bloch point chirality seen in bobbers, opposing domes have oppositely ``charged" points. This is the same charge as the topological charge of a skyrmion since a Bloch point is a skyrmion projected up to small spherical region. Thus, the hourglass-dome transition conserves total charge. On the other hand, what is not conserved is the skyrmion charge in the planes between the opposing domes. The skyrmion creation/annihilation energy barrier in the discrete limit is overcome by the transition. This scenario is in contrast to the decay of skyrmion tubes in chiral magnets into bobbers, where the elimination of the tube results in the appearance of two repulsive bobbers.

The interactions between domes can be described by an effective interaction between Bloch points. To do this, consider that at very large film thickness $H$, the energy is independent of thickness. This constant energy is that of the two domes $E_{2D}$. For the present discussion, we limit the detailed description of this quantity, but elaborate in the Supplemental Material. $\Delta$ sets the length scale for the system, so as $H$ decreases adiabatically and approaches twice the domes' equilibrium height $h$ plus $\Delta$, a force of attraction $E_I(H-2h)$ appears between the Bloch points. We model this short range attraction by a Yukawa potential in the form
\begin{equation}\label{eq:Yukawa}
    E_I = -\mathcal{E}\frac{e^{-2(H-2h)/\Delta}}{H-2h}
\end{equation}
This validity of this model is evident in simulation data shown in Fig.~\ref{fig:Comp_energylandscape}. We also see from the data that the size of the dome, represented by the fixed boundary $R_0$, inversely affects the interaction energy strength $\mathcal{E}$. At close distances, interacting domes will elongate due to this interaction, thereby warping their equilibrium shape.

At the low thickness limit, we have a tube whose energy $E_T(H)$ is monotonically increasing with thickness. Eventually, at the critical height $H_c$ shown in Fig.~\ref{fig:phaseDiagram}, its energy will equal that of the two interacting domes:
\begin{equation}\label{eq:Hc}
    E_T(H_c) = E_{2D} + E_I(H_c - 2h)
\end{equation}
The midpoint thickness of an hourglass decreases inversely to the height of the film, i.e. it gets thinner as it becomes taller. Once this diameter roughly reaches $\Delta$, it pinches closed and splits into the two oppositely charged Bloch points forming domes. 

The complete energy profiles of both systems are shown for a number of parameters in Fig.~\ref{fig:Comp_energylandscape}(c), normalized by their respective two dome energy $E_{2D}$. Most importantly, we see the energy equivalence of the hourglass and double dome solutions clearly at the respective $H_c$ where the curves intersect.

We have show that in ferromagnets without bulk DMI, there are two distinct metastable structures localized in 3D originating from interfacial skyrmions. The equilibrium properties of these structures has been described and their stability over a range of film thicknesses has been presented. They can be thought of as the symmetric analogues to skyrmion tubes and chiral bobbers in noncentrosymmetric ferromagnets, but their properties are surprisingly different. We found that at a certain critical film thickness their energies are equivalent and they may transition reversibly between each other. The interaction between the oppositely charged Bloch points that terminate a pair of opposing domes has been described through an effective interaction model.

The most intriguing finding of this study is the possibility to switch between configurations within a specified range of film thicknesses. This transition offers a unique mechanism for dynamically controlling distinct topological states in three-dimensional ferromagnetic systems. The ability to toggle between these configurations is not only of theoretical interest but also provides a tangible experimental opportunity for controlling topological states in magnetic materials. Such a transition could be exploited to store or transmit information, where external perturbations such as current or magnetic field pulses can trigger state switching. In practical terms, this opens up avenues for reconfigurable magnetic devices. The ability to transition with minimal energy input provides a platform for the development of energy-efficient memory storage and logic devices.

We expect these structures exist around pinning centers at the interfaces of systems that exhibit interfacial DMI such as with Pt/Co/Ir multilayers \cite{Maccariello_2018} or hexagonal Fe films on an Ir(111) surface \cite{heinze_von_2011}. Their 3D aspect means that visualizing their structure could be performed with TEM techniques such as off-axis electron holography \cite{park_2014,Midgley2009} or using the polarization dependence of resonant elastic x-ray scattering (REXS) \cite{van_der_Laan2021}. Further experimental detection could be achieved by soft X-ray magnetic tomography, based on electron holography or X-ray magnetic circular dichroism\cite{Tejo_2021,Hierro-Rodriguez_2020,Donnelly_2020}. The unique structure of Bloch points dominates any emergent magnetic properties, so spontaneous production of a pair of them will significantly alter the associated topological Hall responses\cite{Donnelly_2020_2,Li_2021}.

Nearly all of the numerical calculations in research were conducted with the advanced computing resources provided by Texas A\&M High Performance Research Computing.

\bibliography{domePaper}

\appendix
\input{supp_appendix_arxiv}

\end{document}

%% file: supp_appendix_arxiv.tex
\begin{center}
    \textbf{Supplemental Material for ``Thickness-Driven Transitions Between Novel Magnetic States in Ferromagnetic Films''}\\
    Jacob Mankenberg, Artem Abanov\\
    Department of Physics, Texas A\&M University, College Station, Texas 77843-4242, USA
\end{center}

This supplemental material provides additional details on the methods and calculations used in the main text, including an analytical derivation of the magnetization configurations for the hourglass and dome with numerical solutions, further explanation of the energy landscape, and a deeper discussion of the energies for the equilibrium configurations. We also provide additional numerical results and visualizations.

\vspace{5mm}

Starting with the micromagnetic Hamiltonian $\mathcal{H}$ (Eq.~(1) in the main text) we can use an ansatz to derive equations for the magnetization configurations of both the hourglass and dome solutions. With DMI absent in the bulk but present at the interface we expect a configuration of skyrmions at each layer whose radius diminishes as a function of z. This motivates the following configuration: on a layer at coordinate $z$ we start with the ansatz
\begin{eqnarray}\label{eq:config}
m_{x}=\sqrt{1-s^{2}\left(\frac{r-R(z)}{\Delta}\right)}\cos \phi ,\nonumber\\   
m_{y}=\sqrt{1-s^{2}\left(\frac{r-R(z)}{\Delta}\right)}\sin \phi ,\nonumber\\   
m_{z}=s\left(\frac{r-R(z)}{\Delta}\right),
\end{eqnarray}
where $r$ and $\phi $ are the polar coordinates in the $x-y$ plane, and the function $s$ is monotonically decreasing with the properties
\begin{equation}\label{eq:s}
 s(-R/\Delta)=1,\qquad s(\infty )=-1,\qquad s(0)=0.
\end{equation}
This configuration has a skyrmion charge $1$ on each layer and satisfies the algebraic constraint on the magnetization. The skyrmionic object that forms has  z-dependent radius $R$, i.e. the radius of the curve where $m_{z}=0$. When considering chiral solutions, under the cosine and sine in Eq.~(\ref{eq:config}) one could include the internal angle by writing $\phi +\psi $ where $\psi$  specifically depends on $z$, thus encompassing any stable twisting \cite{mochizuki_2012, onose_2012}. This also distinguishes between N\'{e}el and Bloch type domes \cite{zhou_2015}.

\begin{figure}[ht]
\includegraphics[width=3.5in]{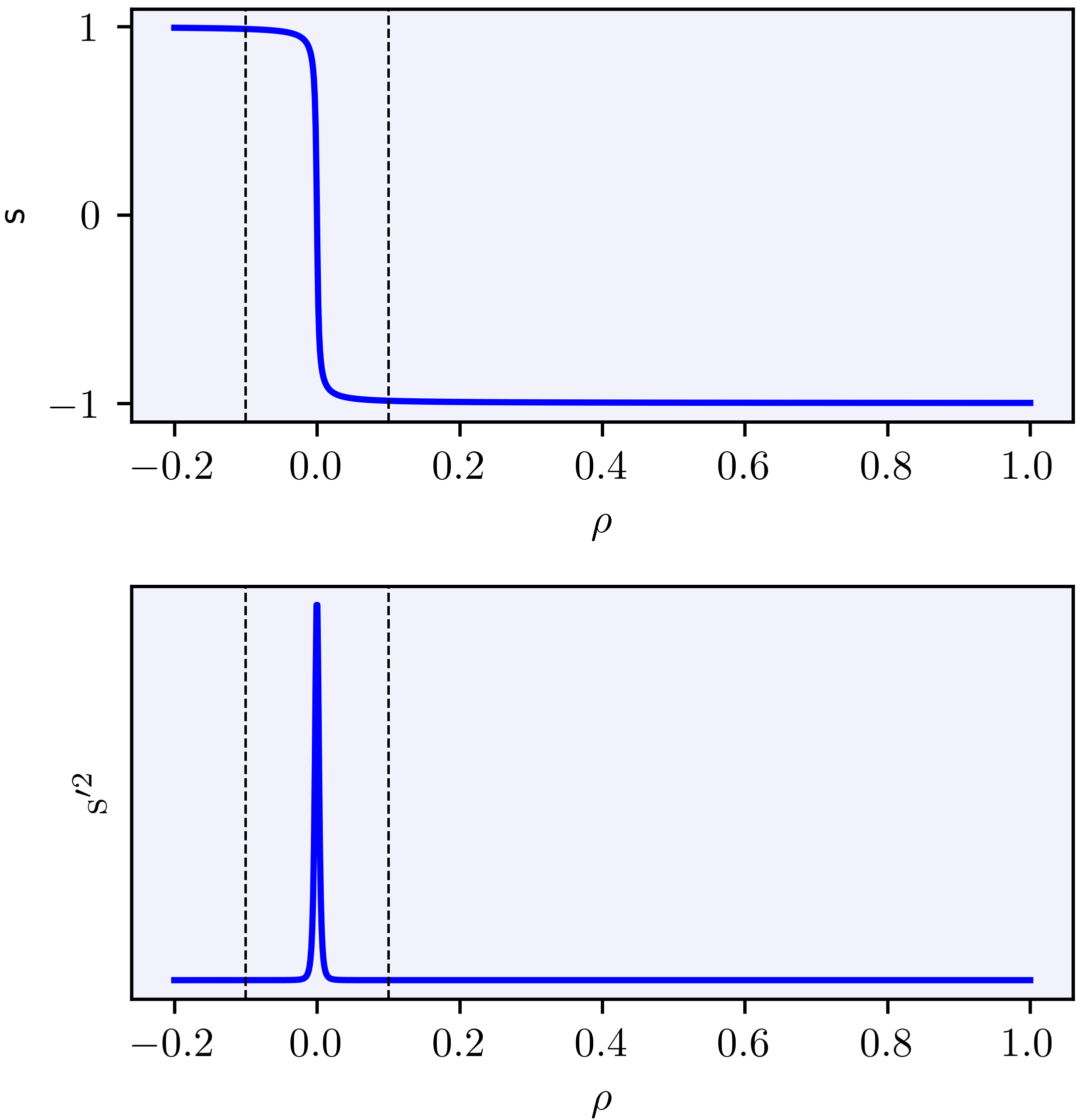}%
\caption{\label{fig:sAndsPrime} Example numerical graph of the profile function $s$ plotted as a function of $\rho = (r - R)/\Delta$. Here, $\Delta = 0.2$ and $R = 1$ so that $R > \Delta$. In this limit, $s$ can be considered odd in the vicinity of $R$, so its first derivative (and any higher powers) are even. This approximation allows us to derive the form of the energy functional in Eq.~(\ref{eq:ERz}).}
\end{figure}

On the very bottom layer interfacial DMI fixes the radius $R(z=0)=R_{0}$. Without this boundary condition, the anisotropy term will collapse the skyrmion on each layer and upon energy minimization we will be left with the trivial phase.

Substituting this ansatz, Eq.~(\ref{eq:config}), into the magnetic Hamiltonian $\mathcal{H}$ we obtain the energy of the dome configuration as a functional of $R(z)$. Consider the radial integration of the z-exchange term explicitly:
\begin{equation}
    J_z\int \text{d}r~r(\partial_zs)^2 = J_z\int\text{d}r~r\left(s'\frac{R}{\Delta}\right)^2.
\end{equation}
With the dimensionless substitution
\begin{equation}
    \rho = \left(\frac{r-R}{\Delta}\right),
\end{equation}
this integral becomes
\begin{equation}
    J_zR'^2\int\frac{\text{d}\rho}{\Delta}(\Delta\rho + R)s'^2.
\end{equation}
Crucially, for $R > \Delta$ the profile $s$ can be considered an odd function. Therefore, $s'^2$ is an even function, and the first term in the integral is zero. We are now left with the second term
\begin{equation}
    J_zRR'^2\int\frac{\text{d}\rho}{\Delta}s'^2.
\end{equation}
where the integral is constant in z. A similar procedure for the other terms in the Hamiltonian gives us the following energy functional
\begin{equation}\label{eq:ERz}
E[R(z)]=\int_{0}^{H}\text{d}z\left(J_{z}BR(\partial_{z}R)^{2} + (\lambda C + JA)R \right),  
\end{equation}
where $H$ is the film thickness and $A$, $B$, and $C$ are $R$ invariant quantities expressed as integrals of $s$ over $\rho$.
\begin{figure*}[ht]
\includegraphics[width=\textwidth]{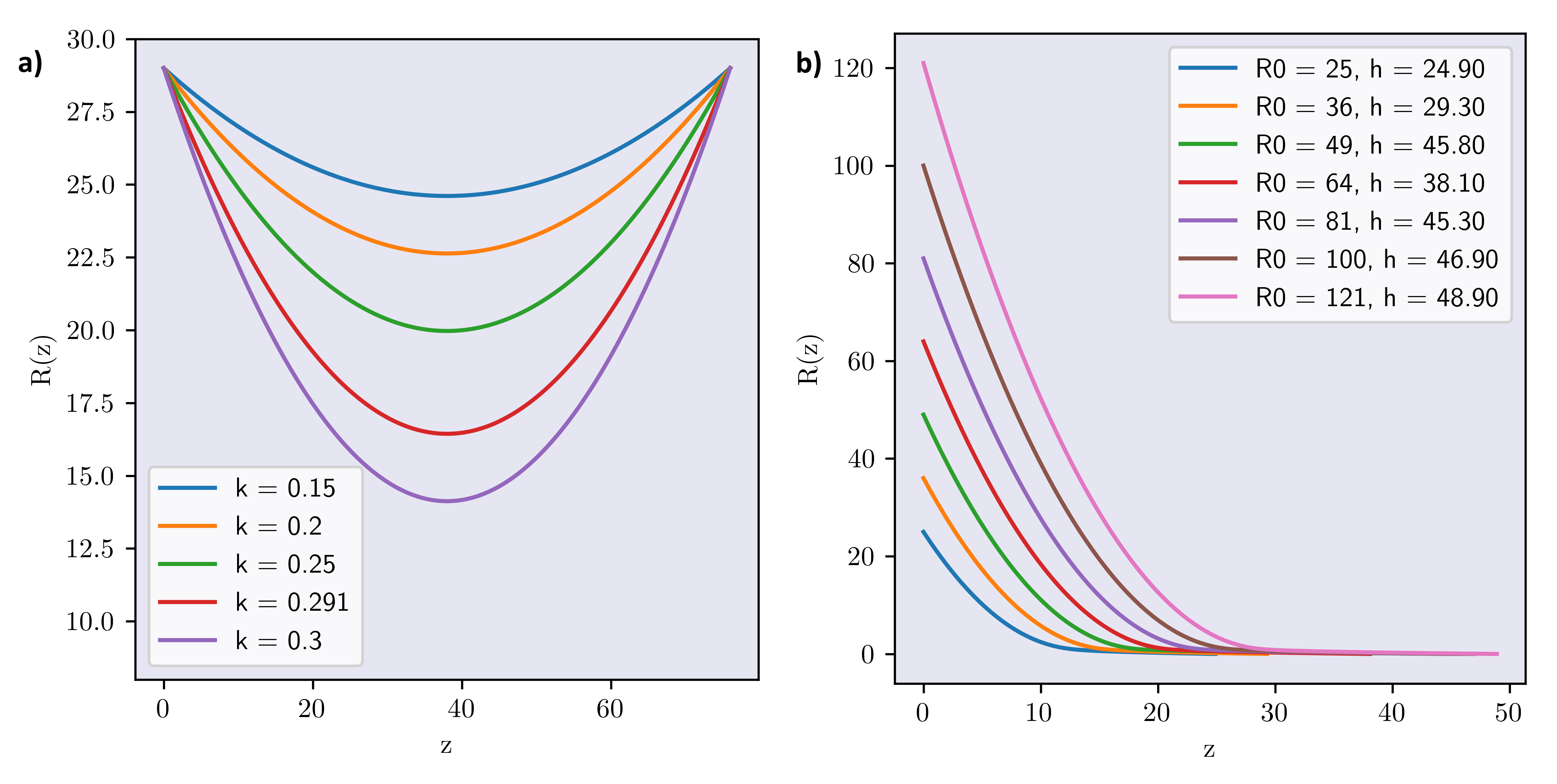}%
\caption{\label{fig:radiiHD} Numerical integration results for the Euler-Lagrange equation Eq.~(\ref{eq:EuLag}). Axes are dimensionless. In a) we have the hourglass solution for a range of values of the material index $k$. We pick the value $k = 0.291$ to match with simulations shown in the main text, such as in the very first image. In b) we impose $R(h) = 0$ conditions, giving the dome solutions, where $h$ is then found by substituting the solution into the energy functional Eq.~(\ref{eq:ERz}) and minimizing for $h$. This is shown for a range of values of $R_0$ and the resulting $h$ is listed in the legend. An obvious inconsistency with real systems is that this continuum model greatly overestimates the equilibrium height $h$, as seen in the tails leading off to the right.}
\end{figure*}

The resulting Euler-Lagrange equation for $R(z)$ is
\begin{equation}\label{eq:EuLag}
    \frac{d}{dz}(RR') = k + \frac{1}{2}R'^2
\end{equation}
where
\begin{equation}
    k = \frac{\lambda C + JA}{2J_zB}
\end{equation}
is a material parameter and the prime denotes derivation by z. The first boundary condition is $R(0) = R_0$ as previously discussed, and the second depends on the structure we are looking for. For a dome, we say that at some unknown height $z = h < H$ we have $R(h) = 0$. Then $h$ is determined by substituting the solution back into the energy, and minimizing: $\partial E/\partial h = 0$. On the other hand, for the hourglass, we simply impose $R(H) = R_0$. The solution for $R$ must be nonnegative everywhere, and this condition sets the limits on $H$. Realistically, when $R \approx \Delta$, the domain wall width, the hourglass solution collapses into the two domes. This determines the critical height $H_c$. 

To solve the equation of motion first note that the energy is translation invariant along z so performing a Legendre transformation and using the first integral we can reduce the differential equation to a separable first order. For notational simplicity, divide Eq.~(\ref{eq:ERz}) by $2J_zB$ and call this reduced energy $\tilde{E}$. Under the Legendre transformation its first integral is
\begin{equation}
    \mathcal{E} \equiv \frac{1}{2}RR'^2-kR.
\end{equation}
Solving for $R'$ and integrating we have
\begin{eqnarray}\label{eq:genSol}
    &z(R) = \int\frac{\text{d}R}{\sqrt{2\mathcal{E}/R + 2k}} \\
    &= \frac{R\sqrt{k^2+\mathcal{E}k/R} - \mathcal{E}\tanh^{-1}(\sqrt{1+\mathcal{E}/kR})}{\sqrt{2k^3}} + C
\end{eqnarray}
Inverting this gives the general solution $R(z)$ and the integration constant $C$ accounts for $R_0$ while for an hourglass, the value of $\mathcal{E}$ is given by $R(H) = R_0$. For domes on the other hand, $\mathcal{E}$ is found by $\partial\tilde{E}/\partial h = 0$:
\begin{eqnarray}
    \frac{\partial\tilde{E}}{\partial h} = \frac{\partial}{\partial h}\int_0^h\text{d}z~(\mathcal{E} + RR'^2) = \\\mathcal{E} +\frac{\partial}{\partial h}\int_0^h\text{d}z~(RR'^2) = 0
\end{eqnarray}
Substituting the general solution Eq.~(\ref{eq:genSol}), evaluating the integral and performing the derivative gives us the form of $\mathcal{E}$ that yields dome solutions.

A numerical integration of the solutions is shown in Fig.~(\ref{fig:radiiHD}). In subplot (a) are the profiles for an hourglass solution for varying values of the material parameter $k$. The physical one corresponding to the systems constituting the bulk of the main work is $k = 0.291$. This is the exact hourglass shown in detail in Fig.(1a) of the main text. On the right in (b) are profiles for a single dome with varying boundary $R_0$ for the physical $k = 0.291$.

It is evident that results grossly overestimate the dome heights. Correctly describing the configurations of the domes requires breaking the continuum limit as the radius becomes smaller than the domain wall width. A full treatment would thereby consist of a lattice theory. Furthermore, a full description the energy requires a more detailed geometric framework regarding domain wall membranes in 3D. Winding number, variable domain walls and higher order terms reminiscent of elasticity theory need to be accounted for. We save this treatment for future work.